%% file: main_acm.tex
\documentclass[sigplan,10pt]{acmart}
\renewcommand\footnotetextcopyrightpermission[1]{}
\usepackage{amsmath, amsfonts, stmaryrd, bbding, makecell}
\usepackage{algorithmic}
\usepackage{graphicx, adjustbox}
\usepackage{textcomp}
\usepackage{xcolor}
\usepackage{comment}
\newcommand{\myname}{\textsc{$\mu$Drive}}

\usepackage{booktabs, multirow}

\usepackage{newfloat}
\DeclareFloatingEnvironment[name=Listing]{listing}
\AtBeginDocument{%
  }

\begin{document}

\title{\textsc{$\mu$Drive}: User-Controlled Autonomous Driving}

\author{Kun Wang}
\affiliation{%
  \institution{Zhejiang University}
  \city{Hangzhou}
  \state{Zhejiang}
  \country{China}
}
\email{kunwang\_yml@zju.edu.cn}

\author{Christopher M. Poskitt}
\affiliation{%
  \institution{Singapore Management University}
  \country{Singapore}
}
\email{cposkitt@smu.edu.sg}

\author{Yang Sun}
\affiliation{%
  \institution{Singapore Management University}
  \country{Singapore}
}
\email{yangsun.2020@phdcs.smu.edu.sg}

\author{Jun Sun}
\affiliation{%
  \institution{Singapore Management University}
  \country{Singapore}
}
\email{junsun@smu.edu.sg}

\author{Jingyi Wang}
\affiliation{%
  \institution{Zhejiang University}
  \city{Hangzhou}
  \state{Zhejiang}
  \country{China}
}
\email{wangjyee@zju.edu.cn}

\author{Peng Cheng}
\affiliation{%
  \institution{Zhejiang University}
  \city{Hangzhou}
  \state{Zhejiang}
  \country{China}
}
\email{lunarheart@zju.edu.cn}

\author{Jiming Chen}
\affiliation{%
  \institution{Zhejiang University}
  \institution{Hangzhou Dianzi University}
  \city{Hangzhou}
  \state{Zhejiang}
  \country{China}
}
\email{cjm@zju.edu.cn}

\renewcommand{\shortauthors}{}

\begin{abstract}
  Autonomous Vehicles~(AVs) rely on sophisticated Autonomous Driving Systems~(ADSs) to provide passengers a satisfying and safe journey.
    The individual preferences of riders plays a crucial role in shaping the perception of safety and comfort while they are in the car.
    Existing ADSs, however, lack mechanisms to systematically capture and integrate rider preferences into their planning modules.
    To bridge this gap, we propose \myname, an event-based Domain-Specific Language~(DSL) designed for specifying autonomous vehicle behaviour.
    \myname~enables users to express their preferences through rules triggered by contextual events, such as encountering obstacles or navigating complex traffic situations.
    These rules dynamically adjust the parameter settings of the ADS planning module, facilitating seamless integration of rider preferences into the driving plan.
    In our evaluation, we demonstrate the feasibility and efficacy of \myname~by integrating it with the Apollo ADS framework.
    Our findings show that users can effectively influence Apollo’s planning through \myname, assisting ADS in achieving improved compliance with traffic regulations. The response time for \myname~commands remains consistently at the second or millisecond level. This suggests that \myname~may help pave the way to more personalizsed and user-centric AV experiences.
\end{abstract}




\settopmatter{printfolios=true}
\maketitle
\pagestyle{plain}

\input{introduction}
\input{background}
\input{language}

\input{evaluation}

\input{relatedwork}
\input{conclusion}

\bibliographystyle{ACM-Reference-Format}
\bibliography{reference2}


\end{document}

%% file: introduction.tex
\section{Introduction}
Autonomous driving systems (ADSs) have undergone significant advancements in recent years \cite{tampuu2020survey,teng2023motion,chen2023end}, integrating sensors and software to control, navigate, and drive autonomous vehicles (AVs). Given the safety-critical nature of ADSs \cite{dixit2016autonomous, favaro2017examining}, it is imperative that they operate safely at all times, even in rare or unexpected scenarios that may not have been explicitly considered during the system's design. This has spurred extensive research into techniques for establishing confidence in an ADS \cite{gu2019towards, bashetty2020deepcrashtest, dosovitskiy2017carla, rong2020lgsvl, li2020av, sun2022lawbreaker}. These methods involve analyzing the ADS before its deployment on actual roads, but achieving coverage for all real-world road scenarios and situations is nearly impossible. In addition, a recent survey conducted by the Institution of Mechanical Engineers \cite{report_2023} sheds light on public perceptions of autonomous vehicles. Two-fifths of respondents expressed their primary concern about traveling in a fully autonomous vehicle, emphasizing the absence of overall human control. Additionally, 70\% of respondents indicated discomfort when traveling in an autonomous vehicle with no human control. Driver discomfort does not solely arise from safety concerns regarding the ADS but is also rooted in the perceived lack of control over the system \cite{report_2023, deruyttere2019talk2car}.

To address these issues, exploring the implementation of a communication channel between the driver and the ADS is a valuable initiative \cite{deruyttere2019talk2car}. This approach proves beneficial not only in alleviating driver discomfort but also in facilitating the effective communication of the driver's preferences to the ADS.
Moreover, in instances where the ADS hesitates or makes errors \cite{report2_2019, sun2024redriver}, the driver can issue commands, actively participating in guiding the vehicle's decision-making process. This active involvement could potentially boost drivers' confidence in AVs, and is particularly crucial for those who lack the ability to manually control the vehicle.
Several existing efforts attempt to integrate the driver as an additional information source to collect data on the driving environment and navigation-related details \cite{deruyttere2019talk2car, kim2019grounding, mirowski2018learning, chen2019touchdown, schumann2021generating, sriram2019talk, pmlr-v100-roh20a, jain2023ground, shah2023lm}. For example, Talk2Car \cite{deruyttere2019talk2car} provides a benchmark for locating reference objects in outdoor environments. The Talk2Nav \cite{mirowski2018learning}, TouchDown \cite{chen2019touchdown}, and Map2Seq \cite{schumann2021generating} introduce tasks of visual language navigation using Google Street View.
However, it is crucial to empower drivers with control over the planning processes of the ADS. At the same time, it is equally important to steer clear of low-level control commands, which is the whole point of autonomous driving in the first place.
Presently, there is no existing method that allows for the direct integration of driver preferences (or intentions) into the planning processes of the ADS.

In this work, we aim to provide a general solution to the driver intent intervention problem for AVs. Specifically, we introduce the \myname~language, an event-driven, domain-specific programming language. This language can seamlessly integrate into state-of-the-art ADSs such as Autoware \cite{Autoware_2024} and Apollo \cite{Apollo_9.0}. \myname~enables drivers to directly intervene in the operational mode of the planning process. It supports driver presets for the driving pattern and real-time issuance of control commands. Drivers can influence the planning module by presetting their driving preferences, such as preferred driving speed, inclination towards lane changes, and permissions for lane borrowing, among other factors. Furthermore, drivers can swiftly modify their driving preferences, planned trajectory, and driving speed in real-time based on the actual driving scenario through instantaneous commands.

\begin{figure}[t]
    \centering
    \includegraphics[width=\linewidth]{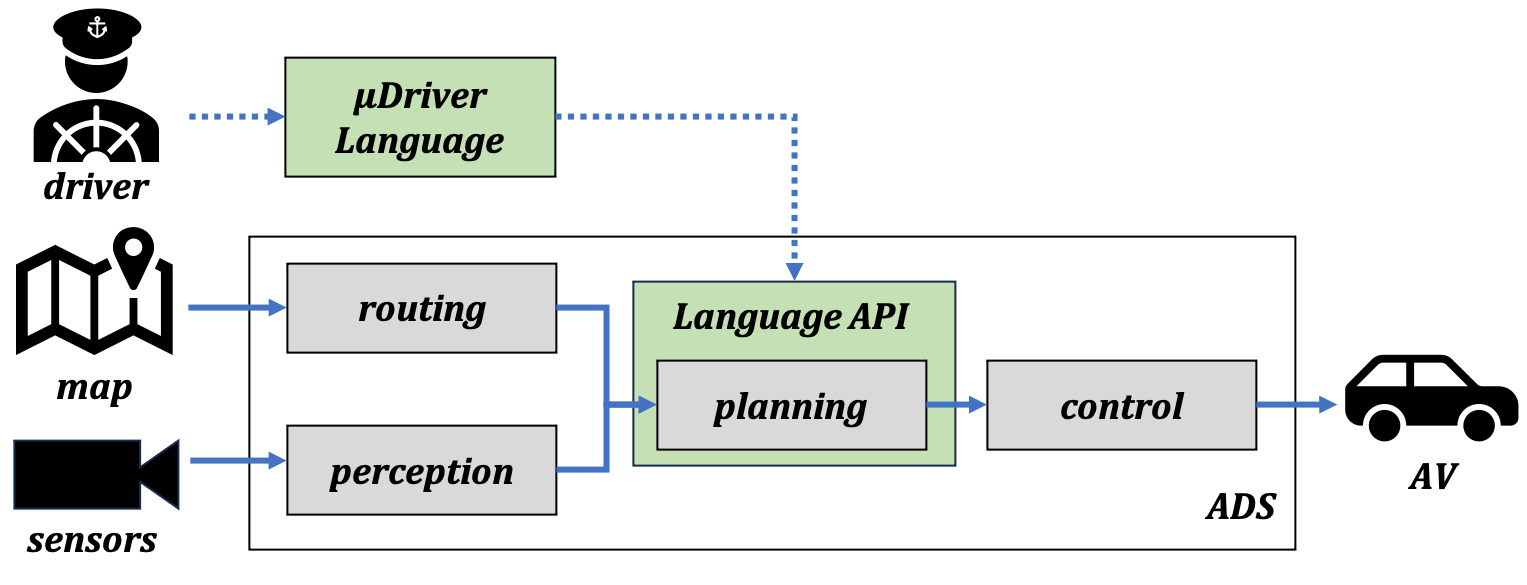}
    \caption{Overflow of \myname}
    \label{fig:overflow}
\end{figure}

\myname~ has been implemented based on the Antlr4 \cite{Antlr_2024} and deployed in the latest Apollo 9.0 \cite{Apollo_9.0}. The implementation comprises two components: a language parser and a language interface. The parser is responsible for parsing driver commands, while the interface ensures that these commands are adhered to during the planning process. This deviates significantly from other works that treat the driver as an auxiliary information source. Existing research primarily utilizes driver information to assist in target detection and destination navigation generation in driving scenarios.

Figure~\ref{fig:overflow} illustrates how \myname~is integrated into the modular design of Apollo. Specifically, two new modules have been introduced, striving to maintain the internal logic integrity of the planning module as much as possible while leaving other modules unchanged. In the diagram, the Routing, Perception, Planning, and Control boxes represent the existing Apollo modules, while the green Language and Interface boxes represent new modules introduced by \myname. Arrows indicate the flow of signal transmission between these modules.

We evaluated our implementation of \myname~against a benchmark of violation-inducing scenarios for Chinese traffic laws \cite{sun2022lawbreaker} and found that users can effectively influence Apollo’s planning, thereby assisting ADS in achieving improved compliance with traffic regulations. Furthermore, the response time for \myname~commands remains consistently at the second or millisecond level.

%% file: background.tex
\section{Background and Problem}
In this section, we review the architecture of the ADS, explore relevant research on leveraging the driver as an auxiliary information source to guide the ADS, and subsequently define the driver intent intervention problem for AVs.

\subsection{Overview of Autonomous Driving Systems}
Traditional ADSs adopt a modular deployment strategy, where functions such as perception, prediction, and planning are individually developed and integrated into the onboard vehicle. State-of-the-art open-source ADSs, exemplified by Apollo \cite{Apollo_9.0} and Autoware \cite{Autoware_2024}, exhibit such architectures. The planning or control module is responsible for generating steering and speed outputs, playing a crucial role in determining the driving experience. The most common planning methods in modular pipelines involve the use of sophisticated rule-based designs.  They are typically structured as loosely coupled modules communicating through message passing.

The routing module within the ADS initially acquires the destination for driving from the high-definition map. Subsequently, it formulates a global route extending from the current vehicle position to the specified destination. Following this, the perception module receives sensor data, such as inputs from cameras or LiDAR sensors. After processing this data, it transmits the refined information to the motion planning module, facilitating the real-time generation of vehicle behavior.
The planning module, in turn, generates a planned trajectory by integrating inputs from the map, route details, feedback obtained from the perception module, and the current state of the ego vehicle—the vehicle under the control of the ADS. Essentially, the planned trajectory delineates the anticipated position of the vehicle at future time intervals. This calculation relies on predictions related to the surrounding environment, encompassing projected trajectories of other vehicles, pedestrians, and responses to traffic signals.
The final control module assumes responsibility for translating the planned trajectory into specific control commands. This encompasses actions related to the throttle, brake, steering wheel, and other relevant components. The primary objective is to ensure that the operational state of the vehicle closely aligns with the predetermined trajectory.

Within a traditional ADS, there may be additional modules, such as the map module found in Apollo, or existing modules could be further segmented into sub-modules, as exemplified by Apollo's breakdown of the planning module into components like the scene module. Despite these variations, the current overarching design of ADS implies the ability to assume control of the existing planning module directly through specific parameters.
This approach allows for the introduction of the driver's driving preferences (or intentions) directly into the planning module without disrupting other modules. For instance, when the driver intends to change lanes, we can easily achieve this by simply modifying the settings of the current scene, without interfering with the functionality of other modules. This way, the system can seamlessly incorporate and respond to driver intent, enhancing the driver experience while maintaining the integrity of the overall ADS architecture.

\subsection{Driver Interaction with ADSs}
In addition to conventional multi-sensor inputs, an increasing number of studies are now considering the driver as an additional source of information to assist ADS in achieving a more stable and secure driving experience \cite{deruyttere2019talk2car,kim2019grounding,mirowski2018learning,chen2019touchdown,schumann2021generating,sriram2019talk,pmlr-v100-roh20a,jain2023ground,shah2023lm}.

Some research endeavors focus on leveraging information provided by drivers as an auxiliary data source to enhance the capability of the perception module. For instance, Thierry et al.~introduced the Talk2Car \cite{deruyttere2019talk2car} dataset with the expectation that AI models can accurately locate reference objects within the driving environment based on the driver's descriptive information. HAD \cite{kim2019grounding} receives driver descriptions of key objects of interest in the current driving environment to influence the focus of the ADS, thereby affecting driving behavior.

Furthermore, a considerable body of research endeavors to utilize driver descriptions as the basis for navigation generation. Specifically, datasets such as Talk2Nav \cite{mirowski2018learning}, TouchDown \cite{chen2019touchdown}, and Map2Seq \cite{schumann2021generating} have introduced tasks involving visual-language navigation using Google Street View. These datasets model the world as a discrete connected graph, requiring navigation to a target in a node-selection format. Sriram et al. \cite{sriram2019talk} encoded natural language commands into high-level behaviors, such as left turn, right turn, or not turning left, and validated their language-guided navigation approach in the CARLA simulator. Junha et al. \cite{pmlr-v100-roh20a} attempted to learn the navigation information embedded in language instructions and used them to adjust a strategy that relies solely on image observations for safe driving of the vehicle. Benefiting from large-scale visual-language pretraining, both CLIP-MC \cite{jain2023ground} and LM-Nav \cite{shah2023lm} leverage CLIP \cite{radford2021learning} to extract language knowledge from instructions and visual features from images. They demonstrate the advantages of pretrained models and present an appealing prototype for addressing complex navigation tasks using multi-modal models.

We distinctly differ from these approaches. Instead of relying on driver information to assist ADS in recognizing road environments and determining focal points to influence ADS operation, we place a stronger emphasis on directly impacting the planning module. Furthermore, these influences are not solely attributed to the road environment; they also encompass some subjective driver preferences and intentions. In addition, we differ from approaches where the driver specifies high-level driving paths. These approaches primarily address situations where maps are unavailable or navigation is not possible. In our case, we still rely on maps and navigation information for driving. Our emphasis is more on assisting the driver in instructing the ADS on the specific driving modes to take on the road. To the best of our knowledge, this work represents the first attempt to directly intervene in the planning module based on driving intent.

\subsection{The Driver Intent Intervention Problem for AVs}
Given an ADS and a user-specified driving demand $\gamma$, our goal is to intervene in the planning mode of ADS at runtime to fulfill the corresponding driving requirements. Addressing this issue can systematically enhance drivers' confidence and adaptability to ADS, thereby increasing their sense of security and satisfaction across various driving scenarios. This intervention strategy not only ensures alignment between ADS planning and execution processes with drivers' expectations but also contributes to optimizing the driving experience. It aims to bring the driving experience closer to users' personalized driving styles and preferences.

The context of this issue presents some crucial requirements for our approach. First, a language with sufficient expressive capability is required to assist drivers in describing their driving intentions. Second, this language should not only support the pre-design of the overall driving pattern but also provide interfaces for real-time adjustments based on road conditions and real-time commands. Third, this language should be high-level so as to free the user from tedious low-level control and yet allow the user to ``control'' the AV effectively at a high-level. Lastly, to ensure its applicability in practice, it must be compatible with existing ADS designs.

%% file: language.tex
\section{The \myname~Language}

We propose \myname, an event-based specification language for autonomous vehicle behaviour.
In this section, we present the syntax of the language, explain how it works via some examples, then introduce its trace-based semantics.

\subsection{Syntax of \myname~Programs}

Figure~\ref{fig:abstract_syntax} presents the abstract syntax of \myname~in EBNF format.
\myname~programs contain one or more rules, each consisting respectively of up to five parts: (1)~a \emph{name} or description expressed as a string; (2)~a \emph{trigger}, which is an event that causes the rule to be applied; (3)~zero or more \emph{conditions}, which constrain the application of rule; (4)~one or more \emph{actions}, which are assignments of driving-related variables that are applied for the duration of the rule; and finally, (5)~an \emph{exit trigger}, which is an event that ends the application of the rule.

\begin{figure}[t]
\begin{center}
\renewcommand{\arraystretch}{1.2}
\begin{tabular}{lcl}
program & ::= & \{rule\}+~ \\
rule & ::= &  '$\mathtt{rule}$' string\_literal \\
  && '$\mathtt{trigger}$' event \\
  && ['$\mathtt{condition}$' \{['$\mathtt{!}$'] condition\}+] \\
  && '$\mathtt{then}$' \{action\}+~ \\
  && ['$\mathtt{until}$' event] \\
  && '$\mathtt{end}$' \\
event & ::= & weather\_event $\mid$ obstacle\_event \\
  && $\mid$ signal\_event $\mid$ road\_event \\
  && $\mid$ '$\mathtt{always}$' \\
condition & ::= & weather\_condition $\mid$ obstacle\_condition \\
  && $\mid$ signal\_condition $\mid$ road\_condition \\
action & ::= & speed\_action $\mid$ distance\_action \\
  && $\mid$ manoeuvre\_action $\mid$ other\_action \\
\end{tabular}
\end{center}
	\caption{Abstract syntax of \myname~programs}
	\label{fig:abstract_syntax}
\end{figure}

We informally describe the behaviour of \myname~programs through two examples.
First, Example~\ref{lst:example_program_vr1} consists of a single rule specifying a driving preference applicable to the island of Madeira, Portugal, where drivers on the VR1 motorway are legally allowed to drive 10km/hr above the normal speed limit in good weather conditions~\cite{reddit_madeira}.
The rule begins to be applied the moment the AV enters a motorway (\verb|entering_motorway|) as long as the weather conditions are good (\verb|!is_raining|, \verb|!is_foggy|, and \verb|!is_snowing|, where \verb|!| indicates negation).
The action applied is to set the AV's maximum speed to 10km/hr above the current value (\verb|increase_max_speed(10)|), which is sustained until the AV leaves the motorway (exit trigger \verb|exiting_motorway|) after which the action is no longer applied.

\begin{listing}[t]
\includegraphics[width=\linewidth]{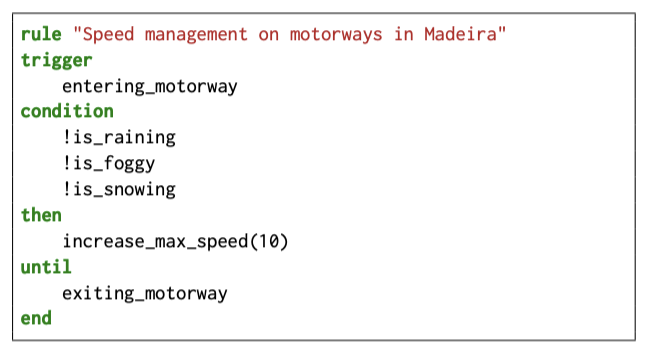}
  \caption{\myname~program specifying that the AV can drive 10km/hr faster on motorways in good weather conditions}
  \label{lst:example_program_vr1}
\end{listing}

Second, Example~\ref{lst:example_program_caution} consists of two rules specifying the behaviour of a cautious AV user at night.
The first rule is triggered upon the detection of another vehicle at night, and causes the AV to switch to low beam and decrease its maximum speed by 5km/hr until the exit trigger event of no vehicles being detected.
The second rule takes over at this point, setting the light to high beam (note also that the maximum speed variable is no longer overridden).
In general, when an event occurs, \myname~applies every rule that has that event as a trigger.
Note that in this example, the triggers, exit triggers, and conditions ensure that the actions of the rules are never in conflict.
For rules with potentially conflicting actions, we describe how \myname~handles them in Section~\ref{sec:language_semantics}.

\begin{listing}[t]
\includegraphics[width=\linewidth]{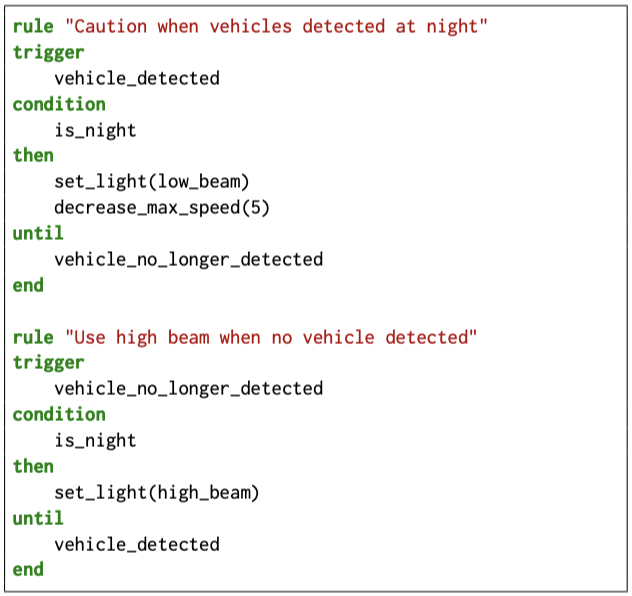}
  \caption{\myname~program specifying that when detecting other vehicles at night, the AV should use low beam and decrease its speed, using high beam and normal speed otherwise}
  \label{lst:example_program_caution}
\end{listing}

\myname~rules are inspired by the structure of IFTTT (if-this-then-that) applets~\cite{ifttt}, which are used to automate digital workflows that integrate multiple different apps and devices.
Similar to IFTTT, \myname~rules are applied upon the occurrence of events subject to certain conditions (`filters' in IFTTT terminology).
However, while IFTTT rules are applied once (e.g.~``send an email if a Tweet mentioning my account is detected''), the actions of \myname~rules are sustained until the occurrence of a given exit trigger.
This sustained application is necessary because events encountered by AV are typically not resolved in a single time-step.
For example, if a driver wishes to be more cautious when encountering an NPC, they would likely wish to do so until the NPC is no longer detected.

\subsection{Triggers, Conditions, and Actions}
\label{sec:triggers_conditions_actions}

We elaborate on the (exit) triggers, conditions, and actions that can be used to construct rules.

\emph{\textbf{Triggers.}} Triggers are based on events that are monitored by \myname~as the AV drives through its environment.
Table~\ref{tab:example_events} lists a few examples of events that can trigger the application of rules.
These are divided into four categories: weather events, which occur when the start or end of various adverse weather conditions are detected; obstacle events, which occur when a nearby vehicle or pedestrian is detected (or no longer detected); signal events, occurring when encountering signals such as traffic lights, stop signs, and speed limit signs; and road events, which correspond to the start and end of various manoeuvres while driving.
Furthermore, the special trigger $\mathtt{always}$ ensures that the rule is applied throughout the entire driving scenario.

\emph{\textbf{Exit Triggers.}} Events may also be used as exit triggers that end the application of a rule.
\myname~provides many events that naturally pair as the start and exit trigger, for example, a rule can be triggered upon detecting a pedestrian until the moment they are no longer nearby.
Specifying an exit trigger, however, is not compulsory.
For example, upon detecting a speed limit sign of 50km/hr, a rule for a cautious driver may assign a maximum speed of 45km/hr, with this action enduring until a later rule overrides this.

\begin{table}[t]
\centering
\caption{Examples of events monitored by \myname}
\footnotesize
\label{tab:example_events}
\begin{tabular}{c|l}
\textbf{Category} & \textbf{Example Events}  \\
\midrule
weather\_event & $\mathtt{rain\_started}$, $\mathtt{rain\_stopped}$, $\mathtt{fog\_started}$, \\
 & $\mathtt{fog\_stopped}$, $\mathtt{snow\_started}$, $\mathtt{snow\_stopped}$   \\
\hline
obstacle\_event & $\mathtt{static\_obstacle\_detected}$, \\ 
 & $\mathtt{pedestrian\_detected}$, $\mathtt{vehicle\_detected}$,  \\
 & $\mathtt{vehicle\_no\_longer\_detected}$ \\
\hline
signal\_event & $\mathtt{red\_light\_detected}$, $\mathtt{green\_light\_detected}$ \\
 & $\mathtt{stop\_sign\_detected}$, $\mathtt{limit(n)\_detected}$ \\
 & $\mathtt{signal\_no\_longer\_detected}$ \\ 
\hline
road\_event & $\mathtt{change\_lane\_started}$, $\mathtt{change\_lane\_finished}$, \\
& $\mathtt{entering\_roundabout}$, $\mathtt{emergency\_stop}$, \\
& $\mathtt{entering\_tunnel}$, $\mathtt{exiting\_tunnel}$ \\
\end{tabular}
\end{table}

\emph{\textbf{Conditions.}} Conditions are used to constrain the application of a rule.
They are analogous to the filters of IFTTT, and are used to specify what must be true of the current environment to allow the rule to be applied.
Table~\ref{tab:example_conditions} lists a few examples of conditions, which are divided into four conditions: conditions about the current detected weather (e.g.~whether it is raining); conditions about obstacles in the vicinity; conditions about current signals such as speed limits or traffic lights; and conditions about the road the AV is currently driving on (e.g.~whether it is a roundabout).
Note that on the occurrence of a rule's trigger, all specified conditions must be simultaneously true for the rule to be applied. It is important to clarify that the specific constraints that can be set heavily depend on the monitoring capabilities of the ADS itself.

\begin{table}[t]
\centering
\caption{Examples of conditions supported by \myname}
\label{tab:example_conditions}
\footnotesize
\begin{tabular}{c|l}
\textbf{Category} & \textbf{Example Events}  \\
\midrule
weather\_condition & $\mathtt{is\_raining}$, $\mathtt{is\_foggy}$, $\mathtt{is\_snowing}$ \\
\hline
obstacle\_condition & $\mathtt{find\_obstacle},$\\
& $\mathtt{obstacle\_distance\_leq(}$number$\mathtt{)}$ \\
\hline
signal\_condition & $\mathtt{find\_signal},$\\
 & $\mathtt{speed\_limit\_geq(}$number$\mathtt{)}$, \\
 & $\mathtt{is\_traffic\_light(}$colour$\mathtt{)}$ \\
\hline
road\_condition & $\mathtt{is\_motorway}$, $\mathtt{is\_roundabout}$, $\mathtt{is\_jam}$ \\
\end{tabular}
\end{table}

\emph{\textbf{Actions.}} Finally, we summarise actions, which can essentially be thought of as variable assignments that are sustained throughout the duration of a rule application.
Actions in \myname~can be grouped into four categories: speed, distance, manoeuvre, and other actions.
They may further be associated with one or more type of triggering event: traffic signals~(S), road scenarios~(R), weather~(W), and dynamic/static obstacles~(O).
Moreover, some actions concern the driver's high-level preferences~(P) and others concern constraints~(C) on the planning module.

First, speed actions allow the driver granular control over speed and acceleration variables.
Table~\ref{tab:speed_action} summarises the actions of this type.
They include actions for setting the default cruising speed, min/max speed, as well as longitudinal/lateral acceleration ranges and reduction ratios.
The driver can also specify a \verb|speed_range(a,b)| action which, while not necessarily having an immediate effect, will ensure that the planning module does not propose actions that bring the AV outside of the speed range $[a,b]$.
Many of these actions are appropriate for obstacle and weather-related rules and triggers, e.g.~allowing a cautious driver to to ensure the AV drives slowly in such scenarios.

\begin{table}[t]
    \centering
    \caption{Speed actions}
    \scriptsize
    \begin{tabular}{c|c|l}
       \textbf{Action} & \textbf{Type} & \textbf{Description}  \\
       \midrule
        \verb|keep_speed(n)| &  & Maintain speed at $n$ km/h, or\\
        && current speed if $n$ is empty.\\
        \verb|max_speed(n)| &  & Set speed $\leq n$ km/h.\\
        \verb|min_speed(n)| &  & Set speed $\geq n$ km/h.\\
        \verb|increase(decrease)| & & Adjust the maximum speed setting.\\
        \verb|_max_speed(n)| & & \\
        \verb|increase(decrease)| & & Adjust the minimum speed setting.\\
        \verb|_min_speed(n)| & & \\
        \verb|increase_to(m,n)| & & Accelerate to speed $n$ km/h with\\
        && acceleration $m$ m/s$^2$; use\\
        && maximum value when $m$ is empty.\\
        \verb|decrease_to(m,n)| & & Decelerate to speed $n$ km/h with\\
        && deceleration $m$ m/s$^2$; use\\
        && maximum value when $m$ is empty.\\
        \verb|cancel_speed_| & & Cancel continuous speed control.\\
        \verb|control|&&\\
        \verb|max_plan_speed(n)| & P & Max speed in planning.\\
        \verb|cruise_speed(n)| & P & Default planning speed.\\
        \verb|near_stop_speed(n)| & P & The speed maintained pre-stop.\\
        \verb|expect_speed(n)| & S \& R & Expected driving speed.\\
        \verb|decrease_ratio(n)| & O \& W & Deceleration speed ratio.\\
        \verb|dec_long_acc_ratio(n)| & W & Longitudinal acceleration reduction\\
        && ration.\\
        \verb|dec_lat_acc_ratio(n)| & W & Lateral acceleration reduction ratio.\\
        \verb|speed_range(n,n)| & C & The speed range used for safety\\
        && checks.\\
        \verb|long_acc_range(n,n)| & C & The longitudinal acceleration range\\
        && used for safety checks.\\
        \verb|lat_acc_range(n,n)| & C & The lateral acceleration range used\\
        && for safety checks.\\
    \end{tabular}
    \label{tab:speed_action}
\end{table}

\begin{figure}[t]
    \centering
    \includegraphics[width=\linewidth]{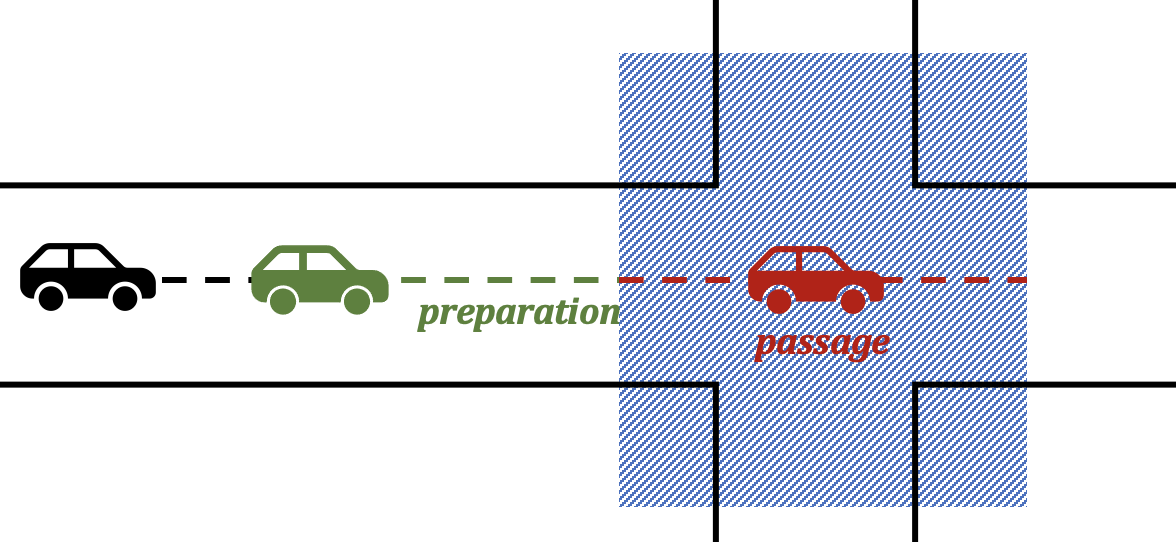}
    \caption{The vehicle passes through special regions.}
    \label{fig:specialarea}
\end{figure}

Second, Table~\ref{tab:distance_action} summarises actions that are used to set variables concerning different aspects of distance.
For instance, the driver is able to configure the longitudinal/lateral buffer distance that the AV should maintain from static obstacles.
When encountering a dynamic obstacle (e.g.~another AV), the \verb|follow_dist(n)| action allows the driver to change the distance they maintain behind it.
In a rule concerning poor weather, for example, a cautious driver may wish to increase it.
Distance actions also allow for quite granular control of how the AV handles special regions such as intersections.
To illustrate, consider Figure~\ref{fig:specialarea}, which depicts the two phases of navigating through them: preparation, followed by passage.
The action \verb|prep_dist(n)| can be used to establish when the first phase begins (a more cautious driver may wish to increase this value).

\begin{table}[t]
    \centering
    \caption{Distance actions}
    \scriptsize
    \begin{tabular}{c|c|l}
       \textbf{Action} & \textbf{Type} & \textbf{Description} \\
       \midrule
       \verb|long_buffer_dist(n)| & O & Longitudinal buffer distance for\\
       && static obstacles.\\
       \verb|lat_buffer_dist(n)| & O & Lateral buffer distance for static\\
       && obstacles.\\
       \verb|follow_dist(n)| & O & Follow distance for dynamic\\
       && obstacles.\\
       \verb|yield_dist(n)| & O & Yield distance for dynamic\\
       && obstacles.\\
       \verb|stop_dist(n)| & O \& S \& R & Min pre-stop distance.\\
       \verb|prep_dist(n)| & S \& R & The preparation distance.\\
       \verb|check_dist(n)| & S \& R & Distance for road inspection.\\
       \verb|expansion_factor(n)| & W & Expansion factor for distance-\\
       && related metrics.\\
    \end{tabular}
    \label{tab:distance_action}
\end{table}

Third, Table~\ref{manoeuvres} summarises actions that correspond to instant manouevres on the road.
For instance, the vehicle can be instructed to change lane, park, pull over, (emergency) stop, or start (if it was already stationary).
The action \verb|lane_follow| instructs the AV to stay in the current lane, which may be useful in the \myname~programs of drivers who prefer smooth and steady journeys to ones with lots of overtaking.

\begin{table}[t]
    \centering
    \caption{Manoeuvre actions}
    \scriptsize
    \begin{tabular}{c|l}
        \textbf{Action}  & \textbf{Description}\\
        \midrule
         \verb|re-planning| & Re-routing and re-planning.\\
         \verb|lane_follow| & Stay in the current lane.\\
         \verb|change_lane(e,n)| & Change lanes to the left (or right) $n$ times.\\
         \verb|park(s)| & Park the vehicle in space $S$.\\
         \verb|pull_over| & Pull over.\\
         \verb|emergency_pull_over| & Emergency pull over.\\
         \verb|stop| & Vehicle stop.\\
         \verb|emergency_stop| & Emergency stop.\\
         \verb|launch| & Start the vehicle when it stops (\verb|pull_over|,\\
         & \verb|stop|, or stop before the intersection).\\
         \verb|cancel_manoeuvre| & Cancel the ongoing effects of commands\\
         \verb|_control| & \verb|lane_follow| and \verb|change_lane|.\\
    \end{tabular}
    \label{behavior}\label{manoeuvres}
\end{table}

Finally, Table~\ref{tab:other_action} summarises a range of other actions that do not quite fit into the previous categories.
A number of them concern the high-level preferences of the driver, e.g.~where to position the car in a lane, whether to use adjacent lanes, or whether to prioritise lane changes.
These are generally specified in rules with the $\mathtt{always}$ trigger so as to maintain these preferences throughout the journey.
For example, (\verb|pri_lane_change(b)|) is helpful for differentiating between a driver who wants drive as efficiently as possible versus one prefers to take it easy.
Some actions involve instructing the car how to behave at certain traffic signs or intersections, e.g.~whether it can turn right on a red light.
Some (e.g.~\verb|comply_signs(b)|) allow the car to excuse itself from certain rules, which may be dangerous in general, but when paired with other actions (e.g.~involving stopping distance and speed) may be helpful in emergency situations.

It should be noted that there may be correlations between different actions, requiring them to be used together. For example, the user-set speed actions \verb|speed_range|, as well as \verb|long_acc_range| and \verb|lat_acc_range| will only take effect when \verb|check_traj| in other actions is set to true. 

\begin{table}[t]
    \centering
    \caption{Other actions}
    \scriptsize
    \begin{tabular}{c|c|l}
         \textbf{Action} & \textbf{Type} & \textbf{Description} \\
         \midrule
         \verb|revise_rule(r,a,v)| & & Adjust the value of action \verb|a| in rule\\
         &&\verb|r| to \verb|v|.\\
         \verb|clear_rule(r)| & & Clean up established rules.\\ 
         \verb|hock_horn| & & Honk the horn.\\
         \verb|set_light(l)| & & Turn on the lights ([high beam, low\\
         && beam, fog light, warning flash]).\\
         \verb|off_light(l)| & & Turn off the lights.\\
         \verb|drive_side(e)| & P & Drive on the left (right, or middle)\\
         && within the lane.\\
         \verb|pri_lane_change(b)| & P & Whether to prioritize lane change.\\
         \verb|borrow_adj_lane(b)| & P & Whether using adjacent lanes.\\
         \verb|obstacle_dec(b)| & O & Whether to decelerate due to obstacles.\\
         \verb|comply_signs(b)| & S & Whether to comply with the traffic sign.\\
         \verb|r_turn_red(b)| & S & Whether right turn on red is permitted.\\
         \verb|time_interval(n)| & R & Time interval between lane changes (s).\\
         \verb|dest_pullover(b)| & R & Whether to pull over when reach\\
         && destination.\\
         \verb|stop_no_sig(b)| & R & Whether to stop at unsignalized\\
         && intersection entry.\\
         \verb|max_hd(n)| & R & Maximum accepted heading deviation.\\
         \verb|max_sp(n)| & R & Maximum accepted steering percentage.\\
         \verb|check_env(b)| & S \& R & Whether to conduct an environmental\\
         && inspection.\\
         \verb|check_speed(b)| & S \& R & Whether to conduct a speed inspection.\\
         \verb|wait_time(n)| & S \& R & Expected waiting time (s).\\
         \verb|crawl(b)| & S \& R & Whether to crawl.\\
         \verb|crawl_time(n)| & S \& R & Expected crawling time (s).\\
         \verb|check_traj(b)| & C & Whether to conduct trajectory checks.
    \end{tabular}
    \label{tab:other_action}
\end{table}

\subsection{Online Rules and Actions}

\myname~programs (i.e.~sets of rules) are generally intended to be written and executed prior to starting a journey.
In reality, however, it is possible that riders will discover on the road that the current program does not fully lead to the driving behaviour they expected.
In these scenarios, the driver should be able to modify or add to the rule set during a journey in order to make it more comfortable.

Besides supporting the addition of more rules, \myname~also supports the idea of `online actions' (or `real-time actions') for situations in which the driver \emph{immediately} wishes to modify a variable, but does not want to wait for a particular event to trigger the corresponding action.
\myname~provides the driver the ability to immediately execute of any of the actions presented in Section~\ref{sec:triggers_conditions_actions} at any point of the journey.

\subsection{Semantics of \myname}
\label{sec:language_semantics}

Intuitively, \myname~operates by monitoring a stream of events that is generated based on the perception module of the ADS.
As well as events, abstract representations of the perceived environment are captured as `scenes', allowing for conditions to be evaluated.
If there are rules whose triggers match any of the events at the current time step, and the conditions are true with respect to the scene, then the actions of the rule are applied.
In particular, this means setting some corresponding parameters of the ADS planning module until the rules no longer apply.

Formally, \myname~can be thought of as a system that monitors a real-time sequence of (abstract) states $\langle S, E \rangle$, and then updates the parameter settings $\Gamma$ of an ADS planning module based on the actions of any activated rules $R$ (initially empty, $\emptyset$).
Here, $S$ denotes an abstract representation of the current \emph{scene} as perceived by the ADS, and is used to evaluate the conditions of \myname~rules; and $E$ is a set of the events that are currently occurring in the ADS's environment.
Note that $S$ and $E$ are constructed by \myname~by interpreting data and messages from the ADS's perception module.
It is worth noting that whether the perception is working correctly is beyond the scope of this work.
At the end of a simulation, these entities together form a trace $\langle \pi_0, \Gamma_0, R_0 \rangle, \langle \pi_1, \Gamma_1, R_1 \rangle, \cdots \langle \pi_n, \Gamma_n, R_n \rangle$, where each $\pi_i = \langle S_i, E_i \rangle$.
Note that the event $\mathtt{always}$ is assumed always to be in each $E_i$.

Let $\mathcal{R}$ denote the set of rules from a \myname~program, each rule $r$ of which is associated with a (triggering) event $e_r$, condition $c_r$, exit event $e'_r$, and set of actions $A_r$.
Given a condition $c_r$, we let $\llbracket c_r \rrbracket_S \in \mathbb{B}$ denote the function that evaluates $c_r$ to a Boolean value.
For instance, $\llbracket \mathtt{is\_night} \rrbracket_S$ evaluates to true if the level of light recorded in the current scene $S$ is below a certain threshold (and false otherwise).
Given an action $a\in A_r$, we let $\llbracket a \rrbracket = \{\mathrm{param}_a \mapsto val_a\}$ for some planning parameter setting $\mathrm{param}_a$ in $\Gamma$ and some value $val_a$ in its domain.
For simplicity, the full definition of $\llbracket \_ \rrbracket$ is left as an implementation matter, as it depends on the specific scene data that can be extracted from an ADS, the implementable planning module parameters, as well as the driver preference thresholds that the \myname~approach is to be tailored for.

We define the semantics of \myname~for the current state $\pi = \langle S, E \rangle$ as follows.
First, a rule $r\in \mathcal{R}$ with triggering event $e_r\in E$ and condition $c_r$ such that $\llbracket c_r \rrbracket_S = \mathrm{true}$ is nondeterministically selected, and the set of activated rules $R$ is updated to $R' = R \cup \{r\}$.
Note that if the actions of $r$ are in conflict with the actions of any rule $r' \in R$, i.e.~the same action exists in both rules but with different arguments (e.g.~\verb|max_speed(30)| and \verb|max_speed(40)|), then $r$ is not added to the set of activated rules.
This can be detected dynamically by tracking the current parameter settings being applied and first checking whether the application of a new rule would impact any of the parameters in that set.
Finally, if the exit trigger $e'_r \in E$, then the rule is removed from the activated rule set $R' = R\setminus \{r\}$.

Second, for every rule $r\in R$, the corresponding actions $A_r$ are applied to the ADS configuration $\Gamma$.
In particular, for each action $a \in A_r$, we update the parameter settings $\Gamma$ of the ADS planning module as $\Gamma' = \Gamma\setminus \{\mathrm{param}_a \mapsto \_ \} \cup \llbracket a \rrbracket$.
Note that the updated settings $\Gamma'$ of the ADS influences the planned trajectories and command sequences it then generates, impacting the scenes and events that are extracted in subsequent states.

Third, for any online actions $A_o$ (initially empty, $\emptyset$) these are applied by updating $\Gamma$ in the way described above.
However, in contrast, if any online action is in conflict with any action set $A_r$ belonging to a rule $r$ in $R$, then $R$ is updated to remove the corresponding rule, i.e.~$R' = R \setminus \{r\}$.
Intuitively, this means that any online actions directly requested by the driver have higher priority than currently activated rules, and the rules are essentially deactivated if they are in conflict.

Ultimately, \myname~can be considered a form of control model that does not require a system model: it boils down simply to setting ADS parameter settings/variables according to the actions of our high-level rules.
In terms of complexity, the number of modifications at each time step is linear in the number of rules.
This is due to our simple semantic model which avoids dependencies between rules, aiming to keep their effects more predictable to the users who write them.

%% file: evaluation.tex
\section{Implementation and Evaluation}
We implement \myname~based on the Antlr4 \cite{Antlr_2024}, and deploy it in the latest Apollo 9.0 \cite{Apollo_9.0}. The code and related results are on our website \cite{UDriver}. Specifically, we designed a language parser to receive \myname~commands, translating them into a series of parameter settings for the Apollo planning module. Subsequently, these parameter-related behaviors are handled through the extended Apollo interface program. 
With these variables, users can effectively take control of the entire planning module. More specifically, these parameter-related behaviors encompass the entire operational cycle of the planning module, involving the parameters utilized, the executed logic, the switching of scenarios, and corrections applied to the planning results. For instance, users can specify the maximum planning speed of the vehicle using \verb|max_plan_speed(n)|, determine whether to pull over upon reaching the destination with \verb|dest_pullover(p)|, and seamlessly transition the vehicle from straight driving to lane changing scenes through \verb|change_lane|. Moreover, \verb|check_trj(b)| empowers users to conduct safety checks on the planned trajectory results. 

We conducted experiments to answer the following Research Questions (RQs):
\begin{itemize}
    \item \textbf{RQ1}: To what extent does the existing ADS system support \myname?
    \item \textbf{RQ2}: Can drivers achieve effective control over AVs through \myname?
    \item \textbf{RQ3}: What is the time interval required from receiving a command to its execution?
\end{itemize}

RQ1 focuses on the level of support for commands in \myname, assessing whether the existing ADS system can fully accommodate the operations specified in \myname.
RQ2 investigates whether \myname~has successfully achieved the primary objectives of exerting control over the ADS. RQ3 delves into the timeliness of control exerted by \myname~on the ADS system, highlighting the runtime execution cost. This assessment is essential for ensuring the effectiveness and efficiency of \myname, as it directly impacts the system's ability to execute commands accurately and promptly, ultimately impacting user experience.

Our experiments are conducted using both Apollo 9.0 and the Apollo Simulation Platform, referred to as Apollo Studio \cite{apollostudio_2024}. Due to the randomness in the simulator, primarily due to concurrency, each experiment is executed 20 times, and we report the average values. All experiments are performed on a machine running Linux (Ubuntu 20.04.5 LTS) equipped with 32GB of memory, an Intel i7-10700k CPU, and an RTX 2080Ti graphics card. 

\textbf{\textit{RQ1: To what extent does the existing ADS system provide support for \myname?}}
In addressing this inquiry, Table \ref{support} delineates the extent of support provided by Apollo for \myname. In the table, the column ``Ours'' represents the extended support of Apollo for \myname. The table uses ``$\times$'' to denote unsupported, ``$\backslash$'' to denote partial support, ``\Checkmark'' for support, and ``-'' to indicate missing. Specifically, the original Apollo does not support drivers to access and configure the ADS's response modes to these events in real-time. For example, once the ADS's default cruising speed is set to 30 km/h, the AV will plan and travel at this preset speed regardless of the actual driving conditions, such as heavy fog or snow, without giving the driver the ability to make adjustments. Therefore, they are classified as unsupported. The extended Apollo, however, supports drivers in specifying response modes for each event specified by \myname. Drivers can use \myname~to set planning parameters such as speed and acceleration for different weather conditions, thereby achieving support for weather events. The missing indicates that the current Apollo does not offer the corresponding interface or implementation. Therefore, the extended version of Apollo also lacks support for the corresponding functionality. 

For constraints, as mentioned earlier, we monitor the results from perception mode to check if the constraint conditions are met. Currently, Apollo provides comprehensive detection for obstacles and traffic signals. Therefore, both Apollo and the extended Apollo offer full support for these constraints. Regarding weather constraints, Apollo does not currently provide detection. Therefore, we have implemented additional interfaces through \myname~to receive weather information, enabling checking of weather conditions. As for road constraints, Apollo only supports some simple road scenario trigger checks, such as determining if it is at an intersection. However, for more complex road information, Apollo's prediction module does not cover scenarios such as whether the upcoming intersection is jammed ($\mathtt{is\_jam}$). Since this work is not involved in perception-related work, we also provide partial support for these constraints. However, it should be noted that for these complex road conditions, the driver can autonomously make decisions and guide the operation of AVs through online rules and actions.

For specific actions, Apollo itself has very limited support because it does not support drivers actively influencing driving behavior. Therefore, we have extensively extended Apollo to help facilitate the effectiveness of driver-initiated actions. Specifically, for behavioral actions, Apollo only supports receiving user commands for emergency pull-over and emergency stop, meaning it performs road maintenance, lane changing, and other operations based on routing results as required. The extended Apollo, however, provides the capability for forced event transitions, such as switching from the lane follow event to the change lane event to facilitate proactive lane changes.
As for distance actions, these commands directly affect distance parameters in the planning process, ensuring the implementation of distance-related behaviors. Regarding speed actions, the extended Apollo adjusts the generated trajectory according to speed commands after the original trajectory generation, ensuring that the speed meets the requirements. Other actions related to vehicle equipment such as lights and horns are temporarily not supported because Apollo currently does not support the relevant interfaces.
It is important to note that, as \myname~is platform-agnostic, some road events (e.g. tunnel and roundabout) and interfaces (e.g. lights and horns) are not implemented in Apollo, leading to partial scene loss. Nevertheless, adapting the extended interfaces to accommodate these scenarios is a straightforward process.

\begin{table}[t]
    \centering
    \caption{The support from Apollo}
    \adjustbox{width=\linewidth}{%
    \begin{tabular}{c|ccc|ccc}
    \toprule
         \multicolumn{2}{c}{Event} & Original & Ours  & Event & Original & Ours\\
         \midrule
         \multirow{3}{*}{Signal} & keep clear & $\times$ & \Checkmark & traffic light & $\times$ & \Checkmark  \\
         & crosswalk & $\times$ & \Checkmark & stop & $\times$ & \Checkmark \\
         & yield & $\times$ & \Checkmark & speed limit & $\times$ & \Checkmark\\
         \midrule
         \multirow{7}{*}{Road} & lane follow & $\times$ & \Checkmark & change lane & $\times$ & \Checkmark\\
         & borrow lane & $\times$ & \Checkmark & reach destination & $\times$ & \Checkmark \\
         & pull over & - & \Checkmark & emergency pull over & $\times$ & \Checkmark \\
         & stop & - & \Checkmark & emergency stop & $\times$ & \Checkmark\\
         & valet park & $\times$ & \Checkmark & open space launch & $\times$ & \Checkmark\\
         & intersection & $\times$ & \Checkmark & others & - & -\\
        \midrule
        \multicolumn{2}{c}{Condition} & Original & Ours & Condition & Original & Ours\\
        \midrule
        \multicolumn{2}{c}{weather} & $\times$ & \Checkmark & obstacle & \Checkmark & \Checkmark \\
        \multicolumn{2}{c}{signal} & \Checkmark & \Checkmark & road & $\backslash$ & $\backslash$\\
        \midrule
        \multicolumn{2}{c}{Action} & Original & Ours & Action & Original & Ours \\
        \midrule
        \multirow{6}{*}{Manoeuvre} & \verb|re-planning| & $\times$ & \Checkmark & \verb|lane_follow| & $\times$ & \Checkmark\\
        & \verb|change_lane| & $\times$ & \Checkmark & \verb|park| & \Checkmark & \Checkmark \\
        & \verb|pull_over| & - & \Checkmark & \verb|emergency_pull_over| & \Checkmark & \Checkmark\\
        & \verb|stop| & - & \Checkmark & \verb|emergency_stop| & \Checkmark & \Checkmark \\
        & \verb|launch| & $\times$ & \Checkmark & \verb|cancel| & - & \Checkmark\\
        \midrule
        \multicolumn{2}{c}{Speed} & $\times$ & \Checkmark & Distance & $\times$ & \Checkmark\\
        \midrule
        \multirow{2}{*}{Other} & \verb|hock_horn| & - & - & \verb|set_light| & - & -\\
        & \verb|off_light| & - & - & others & $\times$ & \Checkmark \\
    \bottomrule
    \end{tabular}}
    \label{support}
\end{table}

\textbf{\textit{RQ2: Can drivers achieve effective control over AVs through \myname?}} To address this issue, we utilize the formalization of traffic regulations reported in Lawbreaker \cite{sun2022lawbreaker} as our specification and assess whether drivers can utilize \myname~to assist ADS compliance with the rules. We note that traffic regulations are quite complex, and the Lawbreaker supports 13 testable traffic regulations and contains numerous sub-clauses. Due to the cessation of maintenance for LGSVL \cite{rong2020lgsvl} and updates to Apollo itself, we have refactored LawBreaker to adapt to Apollo 9.0 and have identified various issues such as Apollo violating traffic regulations, collisions, and inefficiencies through testing. The regulatory descriptions we use can also be accessed on website \cite{UDriver}. Subsequently, we replayed the scenarios that triggered violations, this time with \myname~enabled, to determine whether intervention via \myname~could effectively prevent these violations. 

Specifically, we have incorporated certain official driving practices into Apollo's scenario configurations, drawing from resources such as the California Driver's Handbook \cite{CA2024}, which instructs drivers to ``Be prepared to slow down and stop if necessary'' when nearing an intersection. In alignment with this guidance, we have configured Apollo to approach intersections at speeds not exceeding 15 km/h, as detailed in Example \ref{lst:example_program_intercation}.
Furthermore, we intervene in the Apollo system through online actions to address specific driving conditions that it cannot autonomously monitor, such as detecting congestion at intersections or yielding to through traffic when turning. This intervention enables drivers to respond based on real-time situations, including avoiding congested intersections and stopping to yield when necessary. Example \ref{lst:example_program_stop} provides commands for triggering vehicle stop and vehicle launch in \myname~through online rules and actions. 
Table \ref{useful} illustrates the compliance of AVs with traffic regulations before and after intervention by \myname. Each `sub' in the table represents a sub-rule of the traffic regulation. For example, Law38 pertains to traffic light regulations and comprises three sub-rules, each addressing yellow light, green light, and red light, respectively. The column `Intervention' indicates whether the intervention was successful. The `Improve' column in Table \ref{useful} reports the average improvement of \myname over Apollo.
Please note that due to the close relationship between the sub-rules of Law38 and Law51, they are evaluated together. The sole reason we could not execute certain failed laws is the current lack of support from certain simulators. For instance, we generated a command to turn on fog lights to comply with Law58, but Apollo's vehicle model ignores this command because it currently does not support fog lights. We marked ``Lack support'' in the table to illustrate this point. As shown in Table \ref{useful}, \myname~successfully intervened in all feasible cases. 

\begin{listing}[t]
\includegraphics[width=\linewidth]{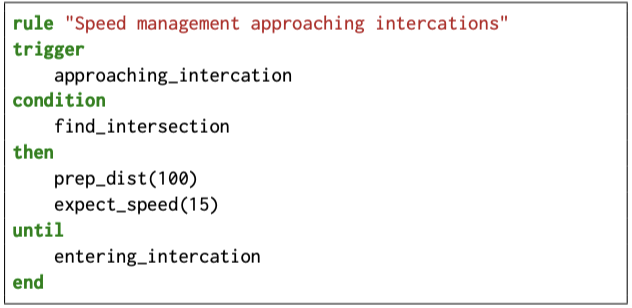}
  \caption{\myname~program specifies that the AV should approach intersections at an expected speed not exceeding 15 km/h when within 100 meters of them}
  \label{lst:example_program_intercation}
\end{listing}

\begin{listing}[t]
\includegraphics[width=\linewidth]{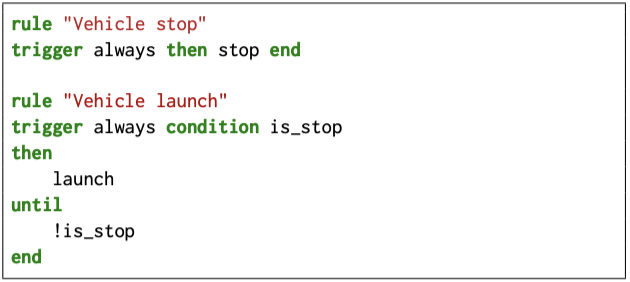}
  \caption{\myname~program specifies the vehicle's stop and launch through online rules and actions}
  \label{lst:example_program_stop}
\end{listing}

\begin{table}[htbp]
    \centering
    \caption{Avoiding violations of Chinese traffic regulations}
    \resizebox{\linewidth}{!}{%
    \begin{tabular}{c|c|c|c|c|c}
    \toprule
       \multicolumn{2}{c|}{Traffic Laws} & Intervention & Improve & Fail Reason & Context\\
    \midrule
       \multirow{3}{*}{Law38} & sub1 & \checkmark & 100\% & - & Green light\\
       & sub2 & \checkmark & 70\% & - & Yellow light \\
       & sub3 & \checkmark & 85\% & - & Red light\\
       \midrule
       \multicolumn{2}{c|}{Law44} & \checkmark & 90\% & - & Lane change\\
       \midrule
       \multirow{2}{*}{Law46} & sub2 & \checkmark & 100\% & - & \multirow{2}{*}{Speed limit}\\
       & sub3  & \checkmark & 100\% & - & \\
       \midrule
       \multicolumn{2}{c|}{Law47} & $\times$ & - & Lack support & Overtake (signals) \\
       \midrule
       \multirow{2}{*}{Law51} & sub3 & $\times$ & - & Lack support & \multirow{3}{*}{\makecell{Intersection with\\traffic lights}}\\
       & sub4 & \checkmark & 100\% & - & \\
       & sub5 & \checkmark & 85\% & - & \\
       \midrule
       \multicolumn{2}{c|}{Law52} & \checkmark & 100\% & - & Without traffic lights\\
       \midrule
       \multicolumn{2}{c|}{Law53} & \checkmark & 100\% & - & Traffic congestion\\
       \midrule
       \multirow{2}{*}{Law57} & sub1 & $\times$ & - & Lack support & Left turn signal\\
       & sub2 & $\times$ & - & Lack support & Right turn signal\\
       \midrule
       \multicolumn{2}{c|}{Law58} & $\times$ & - & Lack support & Warning signal \\
       \midrule
       \multicolumn{2}{c|}{Law59} & $\times$ & - & Lack support & Signals \\
    \bottomrule
    \end{tabular}}
    \label{useful}
\end{table}

To further investigate RQ2, Table \ref{roubutnass} provides a detailed analysis of the compliance of ADS with traffic rules across various scenarios following intervention by \myname. 
The `Pass' column in the table reports the proportion of vehicle compliance with traffic rules. Note that for these scenarios, Apollo's success rate consistently remains below 50\%. However, with intervention through \myname, the AV can completely avoid accidents, violations, and inefficiencies.
The `Robustness' column in the table demonstrates the robustness of AVs in the current scenario to comply with traffic regulations. Specifically, the robustness value represents the distance of the current vehicle trajectory from violating traffic rules. The larger the value, the less likely it is to violate traffic regulations. When it is less than or equal to $0$, it indicates that the corresponding traffic rule has been violated.
In addition, if there are multiple sub-rules within a regulation, the robustness for each sub-rule is sequentially presented. Keep in mind that a lower robustness value signifies a violation is more imminent. For example, in scenario S2, the AV started and entered the intersection when the traffic light was yellow, thereby violating sub2 and sub3 rules outlined in Law38. As a result, the robustness for sub2 and sub3 is 0. After \myname~intervened, although the robustness only increased by 0.5, it signifies that the AV did not violate the corresponding traffic rules.  Symbols `*' and `\#' in the table represent scenarios where the AV is involved in an accident and where the AV fails to reach the designated destination, respectively. Another point worth mentioning is that both scenarios S1 and S9 resulted in traffic accidents due to vehicles not yielding at intersections according to traffic regulations. However, since the LawBreaker module's implementation does not support complex right-of-way checks at intersections, these violations were not identified. However, through manual inspection, it was found that scenario S1 violated Law38\_sub1 (``When the green light is on, vehicles are permitted to proceed, but turning vehicles must not obstruct the passage of straight-moving vehicles or pedestrians''). Scenario S9 violated Law52 (``Turning vehicles yield to straight-moving vehicles''). With the intervention of \myname, Apollo successfully achieved yielding for vehicles with the right of way, resulting in a significant improvement in the robustness of compliance with corresponding regulations. As can be seen from Table \ref{roubutnass}, it is evident that \myname~significantly outperforms the original Apollo in terms of compliance with specifications. Although there may be a slight decrease in robustness, such as in the Sub1 rule of Law38-related scenarios, the values still remain above 0, indicating compliance with the rule. The reason for this decrease in robustness is attributed to our adherence to the requirements outlined in Example \ref{lst:example_program_intercation}, wherein the vehicle decelerates when approaching intersections, ensuring it approaches at a speed not exceeding 15 km/h, thus resulting in a reduction in robustness.

\begin{table*}[]
    \centering
    \caption{Performance comparison of \myname~and Apollo}
    \resizebox{\linewidth}{!}{%
    \begin{tabular}{c|c|c|c|c|l}
    \toprule
       \multicolumn{2}{c|}{Scenario} & \multirow{2}{*}{Driver} & \multirow{2}{*}{Pass} & \multirow{2}{*}{Robustness} & \multicolumn{1}{c}{\multirow{2}{*}{Context}}\\
    \cline{1-2}
    Law & ID & & & \\
    \midrule
    \multirow{10}{*}{\makecell{Law38}} & \multirow{2}{*}{S1*} & Apollo & 0\% & \textcolor{red}{14.4} - 1.0 - 2.0 &The AV entered the intersection during a green light but failed to yield to the straight-moving\\
      \cline{3-5}
      & & \myname & 100\% & \textcolor{green!50!black}{16.0} - 1.0 - 2.0 &  vehicles, resulting in an accident.\\
    \cline{2-6}
    & \multirow{2}{*}{S2} & Apollo & 20\% & 10.7 - \textcolor{red}{0.0} - \textcolor{red}{0.0}  &\multirow{2}{*}{The AV started and entered the intersection when the traffic light was yellow.}\\
    \cline{3-5}
    & & \myname &100\% & 2.0 - \textcolor{green!50!black}{0.5} - \textcolor{green!50!black}{0.5} &\\
    \cline{2-6}
    & \multirow{2}{*}{S3} & Apollo & 40\% & 4.8 - \textcolor{red}{0.0} - \textcolor{red}{0.0}  &\multirow{2}{*}{The AV entered the intersection on a yellow light.}\\
    \cline{3-5}
    & & \myname & 100\% & 1.0 - \textcolor{green!50!black}{0.5} - \textcolor{green!50!black}{0.5} &\\
    \cline{2-6}
    & \multirow{2}{*}{S4} & Apollo & 0\% & 3.1 - 0.3 - \textcolor{red}{0.0} &The AV continued to accelerate and rushed into the intersection on a yellow light,\\
    \cline{3-5}
    & & \myname & 100\% & 2.0 - \textcolor{green!50!black}{0.5} - \textcolor{green!50!black}{0.5} & subsequently passed through the intersection on a red light.\\
    \cline{2-6}
    & \multirow{2}{*}{S5} & Apollo & 0\% & 4.7 - 0.5 - \textcolor{red}{0.0} &\multirow{2}{*}{The AV entered the intersection on a red light.}\\
    \cline{3-5}
    & & \myname & 100\% & 2.0 - 0.5 - \textcolor{green!50!black}{0.5} &\\
    \midrule
    \multirow{4}{*}{Law44} & \multirow{2}{*}{S6} & Apollo & 0\% & \textcolor{red}{-13.8} & The AV is traveling in the fast lane but is not maintaining the required speed limit for the\\
    \cline{3-5}
    & & \myname & 100\% & \textcolor{green!50!black}{6.2} & fast lane.\\
    \cline{2-6}
    & \multirow{2}{*}{S7\#} & Apollo & 20\% & \textcolor{red}{-20.0} & The AV is traveling in the fast lane and come to a stop due to an static obstacle (failure to\\
    \cline{3-5}
    & & \myname & 100\% & \textcolor{green!50!black}{8.8} & change lanes to an available lane on the right), ultimately failing to reach its destination.\\
    \midrule
    \multirow{2}{*}{Law46} & \multirow{2}{*}{S8} & Apollo & 0\% & \textcolor{red}{0.0} - \textcolor{red}{-0.2} & The AV continues to travel at speeds exceeding 30 kilometers per hour despite fog, rain,\\
    \cline{3-5}
    & & \myname & 100\% & \textcolor{green!50!black}{1.2} - \textcolor{green!50!black}{1.2}  & snow, dust storms, and hail.\\
    \midrule
    \multirow{2}{*}{Law52} & \multirow{2}{*}{S9*} & Apollo & 0\% & \textcolor{red}{0.5} & The AV fail to yield to the oncoming straight-through traffic at the stop sign and proceed to\\
    \cline{3-5}
    & & \myname & 100\% & \textcolor{green!50!black}{9.2} & make a left turn at the intersection, resulting in an accident.\\
    \midrule
    \multirow{2}{*}{Law53} & \multirow{2}{*}{S10} & Apollo & 0\% & \textcolor{red}{0.0} & The AV chooses to enter the intersection on a green light despite congestion (6 vehicles\\
    \cline{3-5}
    & & \myname & 100\% & \textcolor{green!50!black}{0.5} & present at the intersection).\\
    \bottomrule
    \end{tabular}}
    \label{roubutnass}
\end{table*}

\begin{figure}[]
    \centering
    \includegraphics[width=1.05\linewidth]{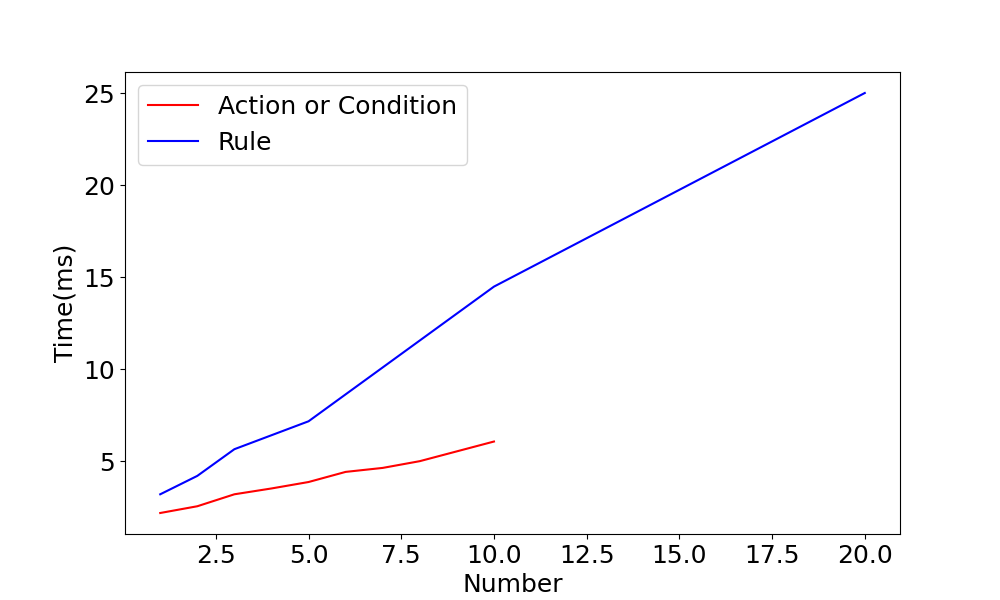}
    \caption{Parsing time of \myname}
    \label{fig:parse}
\end{figure}

\textbf{\textit{RQ3: What is the time interval required from receiving a command to its execution?}} To answer this question, we collect information on the running time of the \myname~for different rules. As illustrated in Figure \ref{fig:overflow}, the operation time of \myname~primarily consists of two parts: parsing driver commands and dispatching specific instructions to Apollo for execution. For parsing driver commands, it operates asynchronously with Apollo itself, thus having no impact on the normal operation of the Apollo system. Figure \ref{fig:parse} illustrates the time required for \myname~to parse driver commands, including the impact of the number of rules and the number of actions and conditions within a rule on parsing time. Overall, the analysis indicates that \myname's parsing duration exhibits a linearly increasing trend with respect to the quantity of rules and the number of actions and conditions per rule. When the number of rules is 1, as the number of actions or conditions within each rule increases from 1 to 10 (in general, the total sum of constraints and actions that can be set within a single rule will not exceed 10), the parsing time of \myname~increases from 2.2 milliseconds to 6.08 milliseconds. On average, it takes 0.98 milliseconds to parse each action or condition, as indicated by the red line in Figure \ref{time}. To investigate the impact of the number of rules on parsing time, we maintained a constant of three actions or conditions per rule. As the number of rules increased from 1 to 20, \myname's parsing time escalated from 3.21 milliseconds to 25.02 milliseconds, averaging a processing rate of one rule every 1.89 milliseconds. Therefore, overall, the parsing time of \myname~remains essentially at the millisecond level.

For time to event, we conduct a statistical analysis based on action types. The detailed data of time to event for \myname~ \\based on Apollo 9.0 is shown in Table \ref{time}. Specifically, if \myname~issues an action, Apollo will process the action from \myname~in the next planning cycle, regardless of whether it involves setting speed and distance parameters or transitioning between various scenarios. Therefore, the time to event of an action heavily depends on the timing of the next planning cycle initiation. In Apollo, once the planning module receives prediction data, it triggers execution by acquiring vehicle chassis and localization information. The official estimate from Apollo is approximately one execution every 100 milliseconds, and the results in the table confirm this observation. Most actions take effect within about 100 milliseconds after being issued. However, it should be noted that the behaviors of re-planning and lane changing consume approximately 1 second each. This is because \verb|Re-planning| action involves rerouting, so there is a wait from the issuance of the action to its actual effect on the planning module while waiting for the routing module's new result. Similarly, the time consumed for lane changing is because Apollo itself does not support active lane changing (as described in RQ1). When extending Apollo to facilitate active lane changing, we rely on the re-planning action. Specifically, this involves adding intermediate waypoints along the target lane, followed by rerouting and planning to accomplish the lane change. Overall, commands issued through \myname~can essentially control the AV in seconds or milliseconds, ensuring the real-time effectiveness of the commands.

\begin{table}[]
    \centering
    \caption{Time to event of action}
    \begin{tabular}{cc|c|c}
    \toprule
       \multicolumn{2}{c|}{\multirow{2}{*}{Action}} & \multicolumn{2}{c}{Time to event}\\
       \cline{3-4}
       & & avg & max\\
    \midrule
       \multicolumn{2}{c|}{{Speed}} & 108.52ms & 135.47ms \\
    \midrule
        \multicolumn{2}{c|}{{Distance}} & 105.36ms & 132.75ms\\
    \midrule
        \multicolumn{1}{c|}{\multirow{3}[3]{*}{Manoeuvre}} & \verb|re-planning| & 954.86ms & 993.20ms\\
        \cmidrule{2-4}
        \multicolumn{1}{c|}{}& \verb|change_lane| & 973.20ms & 1095.37ms\\
        \cmidrule{2-4}
        & \multicolumn{1}{|c|}{\textit{others}} & 113.48ms & 157.46ms\\
        \midrule
        \multicolumn{2}{c|}{Other} & 117.85ms & 153.76ms \\
    \bottomrule
    \end{tabular}
    \label{time}
\end{table}

%% file: relatedwork.tex
\section{Related Work}
The paramount importance of safety in ADS has driven a significant number of research initiatives, concentrating on runtime intervention. These efforts are directed towards bolstering the systems' reliability and elevating users' trust in ADS.

As described before, some initiatives have aimed to integrate linguistic knowledge into driving behaviors. Deruyttere et al. \cite{deruyttere2019talk2car} and Kim et al. \cite{kim2019grounding} endeavor to improve the capability of the perception module by harnessing information provided by drivers. Mirowski et al. \cite{mirowski2018learning} propose an interactive navigation environment that utilizes Google Street View for photographic content and global coverage. Chen et al. \cite{chen2019touchdown} investigate the joint reasoning problem of language and vision through navigation and spatial reasoning tasks. Sriram et al. \cite{sriram2019talk} encode natural language instructions into high-level behaviors including turning left, turning right, not turning left, etc., and verify their language-guided navigation approach in the CARLA simulator. Schumann et al. \cite{schumann2021generating} present a neural model that takes OpenStreetMap representations as input and learns to generate navigation instructions that contain visible and salient landmarks from human natural language instructions. 
Note that these works do not directly involve the planning module of ADS. As a complement to these efforts, \myname~ focuses on the real-time operational status of vehicles, which directly integrates user intent into the planning module to guide the operation of AVs.

There is also a considerable body of work attempting to integrate Large Language Models (LLMs) into ADS systems in academic literature \cite{cui2024survey, jin2023surrealdriver, chen2023driving, mao2023gpt, sharan2023llm, fu2024drive, cui2024drive, wang2023chatgpt, huang2022language,wen2023dilu}. Several studies endeavor to employ LLMs for trajectory planning in ADS. For instance, 
Chen et al. \cite{chen2023driving} present a distinctive object-level multimodal LLM architecture, blending vectorized numeric modalities with a pre-trained LLM to enhance contextual comprehension in driving scenarios. 
Mao et al. \cite{mao2023gpt} propose a straightforward yet potent method capable of converting the OpenAI GPT-3.5 model into a dependable motion planner for autonomous vehicles. 
Sharan et al. \cite{sharan2023llm} investigated the potential of leveraging the commonsense reasoning abilities of LLMs such as GPT-4 and Llama2 to generate plans for autonomous driving vehicles.
Fu et al. \cite{fu2024drive} explored the potential of using LLMs to comprehend driving environments in a human-like manner, and analyzed their ability to reason, explain, and memorize when faced with complex scenarios. 
Cui et al. \cite{cui2024drive} propose a novel framework that leverages LLMs to enhance the decision-making process of autonomous driving vehicles. \cite{cui2024drive} proposes a similar concept to ours, which involves accepting the driver's driving intent to guide vehicle operation. However, it is important to note that these methods all propose replacing the existing planning module of ADSs with LLMs, whereas \myname~provides the capability to accept driver intent for existing rule-based planning modules. 

Some research endeavors utilize LLMs as a bridge to facilitate human-machine interaction \cite{wang2023chatgpt}, making them the most closely related to our work in this domain. 
Wang et al. \cite{wang2023chatgpt} devise a comprehensive framework that integrates LLMs as a vehicle ``Co-Pilot'' for driving. This framework is capable of executing specific driving tasks while ensuring human intentions are met based on provided information. However, unlike \myname, \cite{wang2023chatgpt} opted for the controller as the actual control object, meaning direct interaction with the control model. Compared to directly intervening in the control module, \myname's intervention in the planning module can provide higher levels of safety and reliability, while better aligning with the overall intent of the driver.

There are some efforts focused on runtime enforcement \cite{sun2024redriver, 9172045, grieser2020assuring, shankar2020formal}. Sun et al. \cite{sun2024redriver} monitor the planning trajectory of ADS based on the quantitative semantics of user-defined properties, such as traffic regulations, expressed in signal temporal logic. They employ gradient-driven algorithms to rectify the trajectory in cases where potential violations of regulations are detected.
Guardauto et al. \cite{9172045} partition the ADS into distinct segments to detect rogue behaviors, subsequently restarting the affected partition to rectify them. Grieser et al. \cite{grieser2020assuring} construct an end-to-end neural network, spanning from LIDAR data to torque/steering commands, which implicitly learns safety rules. Additionally, it continuously monitors the distance to obstacles along the current trajectory and activates emergency brakes if a collision is imminent. Shankaro et al. \cite{shankar2020formal} devise a policy utilizing an automaton and mandate the vehicle to halt in case of policy violation. Generally, while these methods exert an influence on the planning outcomes of ADS, they are primarily based on criteria such as collision avoidance and adherence to traffic regulations, lacking sufficient expressive and intervention capabilities to address user intentions.

There are also some efforts focused on designing domain-specific languages for AVs \cite{althoff2017commonroad, fremont2019scenic, queiroz2019geoscenario, dreossi2019verifai, tuncali2019requirements, censi2019liability, collin2020safety, esterle2020formalizing, maierhofer2020formalization, rizaldi2015formalising, rizaldi2017formalising, 10064002}.
Althoff et al. \cite{althoff2017commonroad} propose composable benchmarks for motion planning on roads (CommonRoad).
Fremont et al. \cite{fremont2019scenic}  propose a new probabilistic programming language for the design and analysis of perception systems, especially those based on machine learning.
Queiroz et al. \cite{queiroz2019geoscenario} propose a language to formally capture test scenarios that cover the complexity of road traffic situations.
Zhou et al.~\cite{10064002} propose AVUnit, a framework for systematically testing AV systems against customizable correctness specifications. Subsequently, Sun et al. propose Lawbreaker \cite{sun2022lawbreaker}, an automated framework for testing ADSs against real-world traffic laws, which is designed to be compatible with different scenario description languages. Censi et al. \cite{censi2019liability} and Collin et al. \cite{collin2020safety} describe traffic laws by connecting atomic rules. The works \cite{esterle2020formalizing, maierhofer2020formalization, rizaldi2015formalising, rizaldi2017formalising} propose various methods for describing traffic laws. These works either focus on generating test scenarios or are used to describe certain driving rules as test oracles. \myname~focuses on aiding ADS to achieve safer, more stable, and more comfortable driving through driver intervention, which provides an additional layer of assurance for vehicles operating in real-world scenarios.

%% file: conclusion.tex
\section{Conclusion}
In this paper, we proposed, \myname, an event-based specification language for autonomous vehicle behavior. \myname~supports the execution of complex user-provided driving behaviors, such as driving manuals and traffic regulations, in a manner similar to experienced human drivers. Furthermore, \myname~is platform-agnostic, allowing it to integrate with various state-of-the-art ADSs. We implemented and evaluated \myname~within the Apollo ADS and Apollo Studio. Our findings show that users can effectively influence Apollo’s planning through \myname, assisting ADS in achieving improved compliance with traffic regulations.
For complex Chinese traffic regulations, \myname~can enhance Apollo's performance to 100\% with appropriate interventions, while maintaining acceptable performance overhead.

There are several interesting avenues for future work. First, we are interested in attempting to automatically generate \myname~based on the driver's manual using LLM. Furthermore, we are interested in synthesizing fully customized \myname~rules based on the user's driving data, enhancing the adaptability and personalization of the system.